\documentclass[onecolumn,aps,prd,preprintnumbers,showpacs,superscriptaddress,nofootinbib,amsmath,amssymb,floats,floatfix,showkeys,notitlepage]{revtex4}

\usepackage{graphics,epsfig}
\usepackage{epstopdf}
\usepackage{subfigure}
\usepackage{palatino}
\usepackage{changes}
\usepackage[colorlinks=true,linkcolor=blue,urlcolor=blue,citecolor=blue]{hyperref}
\usepackage[toc,page]{appendix}
\usepackage[normalem]{ulem}
\usepackage{adjustbox}
\usepackage{latexsym}
\usepackage{amsmath}
\usepackage{amssymb}
\usepackage{amsfonts}
\usepackage{times}
\usepackage{graphicx}
\usepackage{dcolumn}
\usepackage{amsmath}
\usepackage{array}
\usepackage{wrapfig}
\usepackage{multirow}
\usepackage{tabularx}
\begin{document}

\title{Weak Gravitational Lensing in Dark Matter and Plasma Mediums for Wormhole-like Static Aether Solution}

\author{Wajiha Javed}
\email{wajiha.javed@ue.edu.pk} 
\affiliation{Department of Mathematics, Division of Science and Technology, University of Education, Lahore-54590, Pakistan}

\author{Sibgha Riaz}
\email{sibghariaz993@gmail.com} 
\affiliation{Department of Mathematics, Division of Science and Technology, University of Education, Lahore-54590, Pakistan}

\author{Reggie C. Pantig}
\email{reggie.pantig@dlsu.edu.ph}
\affiliation{Physics Department, De La Salle University, 2401 Taft Avenue, Manila, 1004 Philippines}
\affiliation{Physics Department, Map\'ua University, 658 Muralla St., Intramuros, Manila 1002, Philippines}

\author{Ali {\"O}vg{\"u}n}
\email{ali.ovgun@emu.edu.tr}
\affiliation{Physics Department, Eastern Mediterranean University, Famagusta, 99628 North
Cyprus via Mersin 10, Turkey.}

\begin{abstract}
In this paper, we study the deflection angle for wormhole-like static aether solution by using Gibbons and Werner technique in non-plasma, plasma, and dark matter mediums. For this purpose, we use optical spacetime geometry to calculate the Gaussian optical curvature, then implement the Gauss-Bonnet theorem in weak field limits. Moreover, we compute the deflection angle by using a technique known as Keeton and Petters technique. Furthermore, we analyze the graphical behavior of the bending angle $\psi$ with respect to the impact parameter $b$, mass $m$ as an integration constant, and parameter $q$ in non-plasma and plasma mediums. We examine that the deflection angle is exponentially increasing as direct with charge. Also, we observe that for small values of $b$, $\psi$ increases, and for large values of $b$ the angle decreases. We also considered analysis to the shadow cast of the wormhole relative to an observer at various locations. Comparing it the Schwarzschild shadow, shadow cast is possible for wormhole as $r<2m$. At $r>2m$, the Schwarzschild is larger. As $r\to \infty$, we have seen that the behavior of the shadow, as well as the weak deflection angle, approaches that of the Schwarzschild black hole. Overall, the effect of plasma tends to decrease the value of the observables due to the wormhole geometry.
\end{abstract}
  

\date{\today}
\maketitle

  \section{Introduction}

Einstein's theory of general relativity (GR) looked into the physical existence of black holes (BHs) in the universe \cite{einstein1916approximative:1916xnr}. The American astronomer John Wheeler invented the word `` BH ''. 
The study of BHs have received a lot of attention since the Event Horizon Telescope obtained the first images of the Messier $87$ BH and then Sagittarius A* BH \cite{EventHorizonTelescope:2019dse,EventHorizonTelescope:2022xnr}. Black hole emits Hawking radiation as a complete thermal spectrum by incorporating quantum theory. Stellar, intermediate, super-massive and microscopic BHs are four main types of BHs. The outer horizon, inner event horizon as well as the singularity are the three ``layers'' of a BH. A BH's event horizon is the boundary around a BH and beyond which light cannot escape. Singularity, is a region in space where the density of the existing mass is infinite.

Wormholes (WHs), just like BHs, can be expressed as solution of Einstein field equations. Schwarzschild BH solution is the simplest solution of Einstein field equations. Flamm first proposed the concept of a WH, after the discovery of Schwarzschild's BH solution. A WH is a hypothetical spacetime that connects two separate regions of universe and give a shortcut through them. Einstein and Rosen \cite{einstein1935particle:1935xnr} proposed the existence of WH-like objects, often known as Einstein-Rosen bridges. Misner and Wheeler \cite{wheeler1957nature:1957xnr} formulated the concept of a ``WH'' \cite{misner1957classical:1957xnr}. Wormholes have not been yet physically demonstrated. After that, Wheeler \cite{wheeler1955geons:1955xnr} explained that WHs are unstable and non-traversable even by a photon. Morris and Thorne \cite{morris1988wormholes:1988xnr} invented the term traversable WH. However, Morris, Thorne and Yurtsever \cite{morris1988wormholes:1988xnr} explained how to convert a WH through traversing space into traversing-time. They demonstrated that by computing the Einstein field equations, we get the solution showing WH-geometry in a terms of a static spherically symmetric line-element. After that, by following the Morris-Thorne papers, a lot of physicists looked into WHs from a different point of views \cite{damour2007wormholes:2007xnr,bueno2018echoes:2018xnr,hawking1988wormholes:1988xnr,Ovgun:2018xys,Halilsoy:2013iza,Ovgun:2015sqa}. Later on, another form of traversable WH were introduced by Matt Visser \cite{visser1995lorentzian:1955xnr}, that is known as thin-shell WH in which the path through the WH can be formed in such a way that traversing path does not cross the region of exotic matter. Although, exotic matter causes the problem to create a stable WH. Recently, it has been explained that WHs also play an important part in explaining quantum entanglement \cite{marolf2013gauge:2013xnr}.

The concept of gravitational lensing (GL) due to its gravitational effects occurs when a huge object distorts the space around it and twisting the direction of light passes through it. Gravitational lensing is a strong astrophysical tool for determining the mass of galaxies and clusters, as well as detecting dark matter (DM) \cite{massey2010dark:2010xnr}. There are three types of GL; strong GL, weak GL and micro GL \cite{bartelmann2001weak:2001xnr,cunha2015shadows:2015xnr}. The strong GL enables us to compute the area and intensity of a BH. Moreover, the impact of ``weak GL'' is actually weaker but yet observable analytically. Micro GL isn't the same as strong and weak GL. In this kind of lensing, the lens is small in comparison to weak and strong GL.

Gibbons and Werner \cite{gibbons2008applications:2008xnr} proposed the method to calculate the angle of deflection by various BHs in the weak field limits. The bending angle $\psi$ can be computed by using asymptotically flat spacetime \cite{gibbons2008applications:2008xnr} such as:
\begin{equation}
    \psi=-\int \int_{p_{\infty}} \mathcal{K}dS. \label{sib1}
\end{equation}
Here, $\mathcal{K}$ is the Gaussian curvature, $dS$ is the surface component and ${p_{\infty}}$ stands for infinite domain of the space.

Numerous writers have used GBT to examine the angle of deflection for various BHs and WHs \cite{gibbons2009universal:2009xnr10,gibbons2009stationary:2009xnr9,gibbons2012application:2012xnr8,bloomer2011optical:2011xnr7,werner2012gravitational:2012xnr6,ovgun99weak:2018xnr5,javed2019effect:2009xnr4,javed2019effect:2019xnr3,javed2019effect:2019xnr1,Pantig2020b,Pantig:2020odu,PANTIG2022168722,Pantig:2022toh,Pantig:2022whj,Kuang:2022xjp,Uniyal:2022vdu, Kumaran:2019qqp,Kumaran:2021rgj,sym14102054}
 Javed et al. \cite{javed2020weak:2020xnr1,Javed:2021arr} calculated the weak GL by stringy BH and tidal charged BH. He and Lin \cite{he2016gravitational:2016xnr} examined the GL for Kerr-Newman BH having arbitrary uniform velocity. Crisnejo and Gallo \cite{crisnejo2018weak:2018xnr} looked into the deflection angle of light in the existence of plasma medium. Nakajima and Asada \cite{nakajima2012deflection:2012xnr} studied GL by Ellis WH. Deflection angle for static and axisymmetric rotating Teo WH was examined by Jusufi and Ovgun \cite{jusufi2018gravitational:2018xnr}. Ovgun \cite{ovgun2018light:2018xnr} worked on the light deflection by Damour-Solodukin WHs using GBT.

The discovery of DM \cite{epps2017weak:2017xnr} by weak deflection is an important topic, as it can assist us to understand the massive structure of the universe \cite{bartelmann2001weak:2001xnr}. Zwicky was the $1$st astronomer who proposed the DM. Dark matter is a type of matter which cannot visualize directly. It does not release any light or energy, that's why standard instrument and detectors cannot detect it. Dark Matter consists of  $27\%$ of the total mass energy of the universe \cite{hinshaw2013nine:2013xnr}. Dark matter can be detected by gravitational interaction and  possesses electromagnetic interactions \cite{latimer2013dispersive:2013xnr}. Super-interacting massive particles, weakly interacting massive particles, sterile neutrinos and axions are the types of dark matter candidates. Refractive index used in dark matter maintains the propagation speed. The DM medium's refractive index is defined as \cite{latimer2013dispersive:2013xnr}:
\begin{equation}
n(\omega)=1+\beta{A_0}+{A_2}{\omega^2}.\label{sib8}
\end{equation}
It's important to remember that $\beta=\frac{\rho_o}{4m^2\omega^2}$, ${\rho_o}$ is the mass density of scattered DM particles, $A_o=-2\epsilon^2e^2$ and $A_{2j}\geq0$. The polarizability of the DM candidate is connected to $\mathcal{O}\left({\omega^2}\right)$  and higher terms. The charged DM candidate has an order of $\omega^{-2}$, while the neutral DM candidate has an order of $\omega^2$. Furthermore, if the parity and charge parity inequalities are present, then the linear term appears in $\omega$.

Oost, Mukohyama and Wang \cite{oost2021spherically:2021xnr} obtained the exact stable solution in Einstein-aether theory. The solution is asymptotically smooth, expressed in the isotropic coordinates and specified by two parameters: mass $m$ is an integration constant and $c_{14}$ is a combined coupling parameter. For $c_{14} = 2$, metric reduces to Schwarzschild solution in Einstein theory in isotropic coordinates and for $c_{14} \neq 2$, the solution illustrates  finite size throat that is slightly trapped but smoothly connects the two untrapped patches: one of the patch has a singularity at finite proper distance and other patch is asymptotically flat at infinite proper distance. The aether configuration and spacetime geometry are similar to static WH aether solutions \cite{penrose1969gravitational:1969xnr}.~The WH-like static aether solution is physically undesirable, according to the cosmic censorship conjecture \cite{eling2006spherical:2006xnr}. Moreover, Zhu et al. studied the shadows and deflection angle of charged and slowly rotating black holes in Einstein-Æther theory \cite{Zhu:2019ura}.

One of the goal of this paper is to find the bending angle for WH-like static aether solution by using GBT and Keeton and Petters method. Moreover, we will study the impact of plasma, non-plasma and DM mediums on the deflection angle of given WH. After that, we analyze graphically, the behaviour of $q$, $b$ and $m$ on the bending angle $\psi$. In addition to these goals, while exploring the effects of plasma, we also examined the behavior of the shadow radius of the wormhole based on the different observer locations. To this aim, we have used the methods pioneered by \cite{Perlick2015}, which not only been applied to BH shadows, but as well as wormholes. Since then, the interest regarding the shadow of wormholes have risen \cite{Guerrero:2022qkh,zhu2021shadow:2021xnr,Rahaman:2021web,Bouhmadi-Lopez:2021zwt,Bugaev:2021dna,Peng:2021osd,Guerrero:2021pxt,Wielgus:2020uqz,Wang:2020emr,Gyulchev:2019osj,Ovgun:2018tua,Ovgun:2020gjz,Ovgun:2019jdo,Cimdiker:2021cpz}. 

This paper is organized as follows: in section \textbf{2}, we discuss WH-like static aether solution. In section \textbf{3}, we compute the bending angle of WH in non-plasma medium. Section \textbf{5}, consists of the calculations of the deflection angle in plasma medium. In section \textbf{6}, we discuss the graphical behaviour of the deflection angle in plasma and non-plasma mediums. Section \textbf{7} is based on the study of the effects of DM on the deflection angle. In section \textbf{8}, we find the bending angle by using Keeton and Petters method. In section \textbf{9}, we conclude our results.

\section{Wormhole-like Static Aether Solution} \label{sec2}

The Einstein aether theory is a vector tensor theory which violates Lorentz invariance by connecting the metric to a unit time-like vector field (the aether field) at every point in space. According to the Einstein-aether theory,  WH solutions have been studied. The spherically symmetric WH-like static aether solution of the Einstein field equations connects two separate regions of space and give a shortcut through them.~Exact solution to Einstein-aether theory is exposing analytically in isotropic coordinates $(t, r,\phi, \theta)$, given as follows \cite{zhu2021shadow:2021xnr}:
\begin{equation}
    ds^2=-\left({\frac{1-\frac{m}{2r}}{1+\frac{m}{2r}}}\right)^q dt^2+\frac{({1+\frac{m}{2r}})^{q+2}}{({1-\frac{m}{2r}})^{q-2}}( dr^2+r^2 d\Omega^2), \label{q}
\end{equation}
\begin{equation}
\text{where}~~~~~~~~~~~~~~~~~~~~~~~~~~~~  q=2\left({\frac{2}{2-c_{14}}}\right)^{1/2}\geq2.~~~~~~~~~~~~~~~~~~~~~~~~~~~\nonumber
\end{equation}

Note that $m$ is an integration constant and $c_{14}$ is small nonnegative parameter.
The static spherically symmetric spacetime for WH-like static aether solution can also be written as \cite{zhu2021shadow:2021xnr}
\begin{equation}
~~~~~~~~ds^2=-A(r) dt^2+B(r) dr^2+D(r) d\Omega^2,  \label{AH0}
\end{equation}
\begin{equation}
\text{where}~~~d\Omega^{2}=d\theta^2+\sin^2\theta d\phi^2~~~~~\text{and}\nonumber\\
\end{equation}
\begin{equation}
~~~~A(r)=\left({\frac{1-\frac{m}{2r}}{1+\frac{m}{2r}}}\right)^q,~~~~~~B(r)=\frac{D(r)}{r^2}=\frac{({1+\frac{m}{2r}})^{q+2}}{({1-\frac{m}{2r}})^{q-2}}. \nonumber\\
\end{equation}
Here, mass $m$ is an integration constant, $r$ is radial coordinate and $q$ is parameter. It is noted that $q > 2$, the above metric has a curvature singularity and a marginally trapped throat \cite{Oost:2021tqi}.

\section{Deflection Angle $\psi$ in Non-Plasma Medium} \label{sec3}
In this section, we determine the bending angle of WH in non-plasma medium by using GBT. When source and observer are both in the tropical region and null photon also in the same region, so one can infer that  $(\theta=\frac{\pi}{2})$. In order to get optical metric, we put $ds^{2}=0$ in Eq.($\ref{AH0}$) and get;
\begin{equation}
dt^2=\frac{B(r)}{A(r)}{dr^2}+\frac{B(r)r^2}{A(r)}{d\phi^2}. \label{H1}
\end{equation}
The non-zero Christoffel symbols for the above metric can be obtained as:
\begin{equation}
\Gamma^0_{00}=\frac{1}{2}\left(\frac{-A^\prime(r)}{A(r)}+\frac{B^\prime(r)}{B(r)}\right),~~~~\Gamma^1_{10}=\frac{1}{r}-\frac{A^\prime(r)}{2A(r)}+\frac{B^\prime(r)}{2B(r)},\nonumber
\end{equation}
\begin{equation}
\Gamma^0_{11}=\frac{1}{2}r\left(-2+\frac{rA^\prime(r))}{A(r)}-\frac{rB^\prime(r)}{B(r)}\right),\nonumber
\end{equation}
where $0$ and $1$ are showing the $r$-coordinate and $\phi$-coordinate, respectively and the optical metric's Ricci scalar is calculated as:
\begin{eqnarray}
\mathcal{R}&=&\frac{1}{rA(r)B(r)^3}\left(-rB(r)^2{A^\prime(r)^2}+{A(r)}{B(r)^2}({A^\prime(r)}+{rA^{\prime\prime}(r)})\nonumber
\right.\\&-&\left.{A(r)^2}(-r{B^\prime(r)^2}+{B(r)}({B^\prime(r)}+{rB^{\prime\prime}(r)}))\right).\label{sib3}
\end{eqnarray}
The Gaussian curvature is defined as:
\begin{equation}
\mathcal{K}=\frac{\mathcal{R}}{2}=-\frac{64 m r^3 \left(1-\frac{m}{2 r}\right)^q \left(\frac{m}{2 r}+1\right)^{-q} \left(-\frac{m-2 r}{m+2 r}\right)^q \left(m^2 q-4 m r+4 q r^2\right)}{\left(m^2-4 r^2\right)^4}.\label{sib4}
\end{equation}
For the given WH, using Eq.(\ref{sib3}), the Gaussian curvature is computed as:
\begin{equation}
\mathcal{K}\simeq \frac{-qm}{r^3}+\mathcal{O}\left({m^2}\right).\label{AH1}
\end{equation}
In the region of non-singular domain $\mathcal{H}_{e}$, the deflection angle for WH-like static aether solution by using GBT, can be obtained by using the following formula;
\begin{equation}
\int\int_{\mathcal{H}_{e}}\mathcal{K}dS+\oint_{\partial\mathcal{H}_{e}}kdt
+\sum_{i}\epsilon_{i}=2\pi\xi(\mathcal{H}_{e}),\label{AH6}
\end{equation}
in the above expression, $k$ indicates geodesic curvature, stated as
$k=\bar{g}(\nabla_{\dot{\eta}}\dot{\eta},\ddot{\eta})$ and $\bar{g}
(\dot{\eta},\dot{\eta})=1$, $\ddot{\eta}$ denotes unit acceleration vector
and $\epsilon_{i}$ expresses the exterior angle at the ith vertex.
As $e\rightarrow\infty$, the corresponding jump angles reduce into $\pi/2$ and we
obtain $\theta_{O}+\theta_{S}\rightarrow\pi$. Euler characteristic is
$\xi(\mathcal{H}_{e})=1$. So,
\begin{equation}
\int\int_{\mathcal{H}_{e}}\mathcal{K}dS+\oint_{\partial
\mathcal{H}_{e}}kdt+\epsilon_{i}=2\pi\xi(\mathcal{H}_{e}), \label{sib5}
\end{equation}
here, $\epsilon_{i}=\pi$ represents jump angle. As $e\rightarrow\infty$, the geodesic curvature is obtained as
\begin{equation}
k(D_{e})=\mid\nabla_{\dot{D}_{e}}\dot{D}_{e}\mid.
\end{equation}
Since the radial component of geodesic curvature is;
\begin{equation}
(\nabla_{\dot{D}_{e}}\dot{D}_{e})^{r}=\dot{D}^{\phi}_{e}
\partial_{\phi}\dot{D}^{r}_{e}+\Gamma^{0}_{11}(\dot{D}^{\phi}_{e})^{2}.\label{AH5}
\end{equation}
For large value of $e$, $D_{e}:=r(\phi)=e=const$, then the result is;
\begin{equation}
(\nabla_{\dot{D}^{r}_{e}}\dot{D}^{r}_{e})^{r}\rightarrow\frac{1}{e}.\label{sib6}
\end{equation}
The geodesic curvature does not have a topological defect so, $k(D_{e})\rightarrow e^{-1}$.
However, by using the optical metric Eq.(\ref{H1}), it can be expressed as follows: $dt=ed\phi$.
As a result, we have:
\begin{equation}
k(D_{e})dt=d\phi.\label{sib7}
\end{equation}
Now, using previous expression one can obtain the following equation;
\begin{equation}
\int\int_{\mathcal{H}_{e}}\mathcal{K}ds+\oint_{\partial \mathcal{H}_{e}} kdt
\overset{h \rightarrow\infty}{=}\int\int_{T_{\infty}}\mathcal{K}dS+\int^{\pi+\psi}_{0}d\phi.\label{hamza2}
\end{equation}
The $0$th order light ray in the weak field limits is calculated as $r(t)=b/\sin\phi$.
Using Eqs.(\ref{AH6}) and (\ref{hamza2}), the bending angle can be obtained as
\begin{equation}
~~~~~~~~~~~~\psi=-\int^{\pi}_{0}\int^{\infty}_{b/\sin\phi}\mathcal{K}\sqrt{det\bar{g}} ~dr d\phi,\label{AH7}
\end{equation}
where
\begin{equation}
\sqrt{det\bar{g}}= r+{2qm}+\mathcal{O} \left({m^2}\right).\nonumber
\end{equation}
Using the Gaussian curvature upto the leading order terms and angle of deflection is calculated as
\begin{eqnarray}
\psi &\thickapprox& \frac{2mq}{b}+\mathcal{O} \left({m^2}\right).\label{P1}
\end{eqnarray}
The first term of the obtained deflection angle $\psi$ (\ref{P1}) is depending on the first order of mass $m$, $q$ and $b$. While, the higher order terms of $\psi$ are depending upon higher orders of $m$, $q$ and $b$. For the sake of simplicity, we consider the only first order term of the mass $m$. The obtained bending angle in non-plasma medium converts into the deflection angle of Schwarzschild BH after putting $q=2$.

\section{Deflection Angle $\psi$ in Plasma Medium} \label{sec4}

This section is based on the computation of the deflection angle for WH-like static aether solution in plasma medium. The refractive index $n(r)$ for WH-like solution is calculated as
\begin{equation}
n^2\left(r,\omega(r)\right)=1-\frac{\omega_e^2(r)}{\omega_\infty^2(r)}{A(r)},\nonumber
\end{equation}
which can also be represented as:
\begin{equation}
n(r)=\sqrt{{1-\frac{\omega_e^2}{\omega_\infty^2}\left(A(r)\right)}},
\end{equation}
where electron plasma frequency is denoted by $\omega_{e}$, while $\omega_{\infty}$ denotes photon frequency calculated at infinity by observer, then the corresponding optical metric can be defined as;
\begin{equation}
dt^2=g^{opt}_{lm}dx^ldx^m=n^2 \left(\frac{B(r)}{A(r)}{dr^2}+\frac{B(r)r^2}{A(r)}{d\phi^2} \right).
\end{equation}
For our metric, we can write the above values as:
\begin{equation}
A(r)=\left({\frac{1-\frac{m}{2r}}{1+\frac{m}{2r}}}\right)^q,~~~~~B(r)=\frac{D(r)}{r^2}=\frac{({1+\frac{m}{2r}})^{q+2}}{({1-\frac{m}{2r}})^{q-2}}. \nonumber\\
\end{equation}
The Gaussian optical curvature can be defined as:
\begin{eqnarray}
\mathcal{K}=\frac{A''(r)}{2 B(r) n(r)^2}-\frac{A'(r) B'(r)}{4 B(r)^2 n(r)^2}+\frac{A'(r) D'(r)}{4 B(r) D(r) n(r)^2}-\frac{A'(r)^2}{2 A(r) B(r) n(r)^2} \notag \\ +\frac{A(r) B'(r) D'(r)}{4 B(r)^2 D(r) n(r)^2}-\frac{A(r) D''(r)}{2 B(r) D(r) n(r)^2}+\frac{A(r) D'(r)^2}{4 B(r) D(r)^2 n(r)^2}.\label{hamza1}
\end{eqnarray}

Using Eq.(\ref{hamza1}), the Gaussian optical curvature can be obtained as:
\begin{eqnarray}
~~~~~~~~~~~~~~~~~~~~\mathcal{K}\simeq  -\frac{qm}{r^3}-\frac{3qm}{2r^3}\frac{\omega_e^2}{\omega_\infty^2}+\mathcal{O} \left({m^2}\right).
\end{eqnarray}
We compute the bending angle by using the GBT, for this purpose we apply straight line approximation $r=\frac{b}{sin\phi}$ at 0th order and obtain the deflection angle as;
\begin{equation}
\psi=-\int_{0} ^{\pi} \int_\frac{b}{\sin\phi} ^{\infty} \mathcal{K} dS, \label{sum}
\end{equation}
where $dS$=$\sqrt{-g}dr d\phi$ and
\begin{equation}
dS=r-{r}\frac{\omega_e^2}{\omega_\infty^2}
+\left(2q-q\frac{\omega_e^2}{\omega_\infty^2}\right){m}+\mathcal{O}\left({m^2}\right) dr d\phi.
\end{equation}
Using Eq.(\ref{sum}), the deflection angle in plasma medium can be obtained as:
\begin{eqnarray}
\psi &\thickapprox& \frac{2mq}{b}+\frac{mq}{b}\frac{\omega_e^2}{\omega_\infty^2}\label{P2}.
\end{eqnarray}

\begin{figure}
    \includegraphics[width=0.8\textwidth]{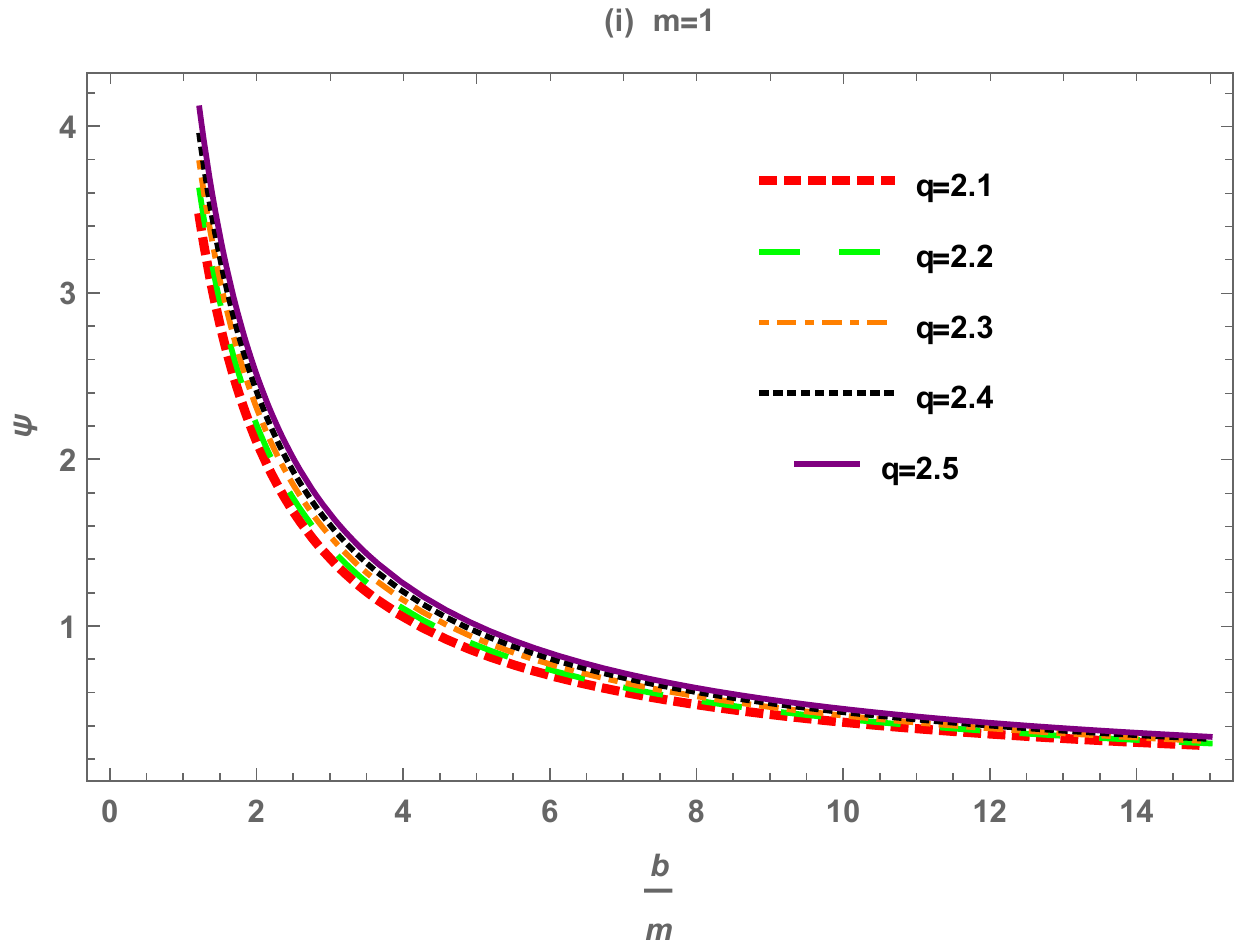}
    \caption{Plot of $\psi$ versus $b$. }
    \label{fig1}
\end{figure}

The obtained deflection angle (\ref{P2}) depends on $m$, $b$ and $q$.~The above results presents that photon rays are travelling into a homogenous plasma medium. We examined that Eq.(\ref{P2}) converted into Eq.(\ref{P1}), if the effect of plasma is neglected. For $q=2$, the obtained angle (\ref{P2}) also reduces into the Schwarzschild angle. 
{ $\psi$ versus $b$}

Here, we use $p=\frac{\omega_e}{\omega_\infty}$=$10^{-1}$ and examine the graphical behaviour of angle $\psi$ w.r.t $b$ and fixed $m=1$ which are discussed below.

\textbf{Figure 1} shows the relationship between $\psi$ and $b$ by varying $q$ and fixing $m$. It consists of small values of $b$ and $m$. When the values of impact parameter $b$ approaches to zero, the deflection angle $\psi$ approaches towards the infinity at $m=1$. As $b$ increases i.e ; $b\rightarrow+\infty$, the $\psi$  approaches to zero. 
It is to be noted that the graphical behaviour of bending angle in the non-plasma case after neglecting the plasma effect will be similar with the plasma case. The impact of the plasma on the bending angle is negligible, one can obtain the same graphs by taking same values in both plasma Eq.(\ref{P2}) and non-plasma mediums Eq.(\ref{P1}).

\section{Deflection Angle in Dark Matter Medium}

This section is devoted to study the impact of the DM medium on deflection angle. For this purpose, we use DM medium's refractive index (\ref{sib8}):
\begin{equation}
n(\omega)=1+\beta{A_0}+{A_2}{\omega^2}.\nonumber
\end{equation}
The WH's optical geometry in 2-dimensional is:
\begin{equation}
~~~~~~~~~~~dt^2=n^2 \left(\frac{B(r)}{A(r)}{dr^2}+\frac{B(r)r^2}{A(r)}{d\phi^2} \right),\label{sib9}
\end{equation}
where
\begin{equation}
A(r)=\left({\frac{1-\frac{m}{2r}}{1+\frac{m}{2r}}}\right)^q,~~~~B(r)=\frac{D(r)}{r^2}=\frac{({1+\frac{m}{2r}})^{q+2}}{({1-\frac{m}{2r}})^{q-2}}.\nonumber
\end{equation}
The Gaussian curvature is expressed as
\begin{eqnarray}
~~~~~~~~~~~~~~~~~~~~~~\mathcal{K}\simeq  -\frac{qm}{r^3(1+\beta{A_0}+{A_2}{\omega^2})^2}+\mathcal{O} \left({m^2}\right).\label{sib11}
\end{eqnarray}
The deflection angle can be obtained as:
\begin{equation}
\psi=-\int_{0} ^{\pi} \int_\frac{b}{\sin\vartheta} ^{\infty} \mathcal{K}\sqrt{det\bar{g}} ~dr d\phi ,\label{sib12}
\end{equation}
For the given line-element one can be determined as;
\begin{equation}
\sqrt{det\bar{g}}=r(1+\beta{A_0}+{A_2}{\omega^2})^2+2qm(1+\beta{A_0}+{A_2}{\omega^2})^2+\mathcal{O}\left({m^2}\right).\label{sib13}
\end{equation}
The deflection angle for WH-like static aether solution in DM medium can be calculated as;
\begin{eqnarray}
\psi &\thickapprox& \frac{2mq}{b(1+\beta{A_0}+{A_2}{\omega^2})^6}, \label{sib14}
\end{eqnarray}
which is depending upon $m$,  $q$ and $b$. If $q=2$, then the obtained bending angle reduces into bending angle of the Schwarzschild BH upto first order of $m$ in dark matter medium. We also observe that the bending angle in case of dark matter medium is larger than in general. This expression simplifies to the vacuum case in the absence of the DM medium.

\section{Deflection Angle by Keeton and Petters Method}

The calculations of the deflection angle $(\psi)$ of the WH-like static aether solution by using the Keeton and Petters technique are discussed in this section.~The PPN (post-post-Newtonian) framework is a direct method to deal with all kinds of gravity ideas for which the weak-deflection limit is stated as in a single variable $m$ in a series expansion. The concept was extended to the $3rd$-order \cite{sarmiento1982parametrized:1982xnr}.~However, Keeton and Petters modified that technique to make it more compatible with their approach and offers new-results \cite{keeton2005formalism::2005xnr}.~The spacetime geometry is supposed to be stable, non-linear spherically symmetric and asymptotically flat:
\begin{eqnarray}
ds^2=-\bar{A}(r)dt^2+\bar{B}(r)dr^2+r^2d\Omega^2.\label{sib16}
\end{eqnarray}
The coefficients of Eq.(\ref{sib16}) should be written in PPN series upto the 3rd-order \cite{keeton2005formalism::2005xnr} as follows:
\begin{eqnarray}
\bar{A}(r)=1+2a_1\left(\frac{\phi}{c^2}\right)+2a_2\left(\frac{\phi}{c^2}\right)^2+2a_3\left(\frac{\phi}{c^2}\right)^3+......\label{sib17}
\end{eqnarray}
\begin{eqnarray}
\bar{B}(r)=1-2b_1\left(\frac{\phi}{c^2}\right)+4b_2\left(\frac{\phi}{c^2}\right)^2-8b_3\left(\frac{\phi}{c^2}\right)^3+......, \label{sib18}
\end{eqnarray}
where $\phi$ is a 3-dimensional Newtonian potential;
\begin{eqnarray}
\frac{\phi}{c^2}=-\frac{m}{r}.\label{sib19}
\end{eqnarray}
The deflection angle in a series form is defined as;
\begin{eqnarray}
\psi=A_1\left(\frac{m}{b}\right)+A_2\left(\frac{m}{b}\right)^2+A_3\left(\frac{m}{b}\right)^3+\mathcal{O}\left(\frac{m}{b}\right)^4, \label{sib20}
\end{eqnarray}
where
\begin{eqnarray}
A_{1} &=& 2(a_{1}+b_{1}),\nonumber
\\A_{2}& =& \left(2a_{1}^2-a_{2}+a_{1}b_{1}-\frac{b_{1}^2}{4}+b_{2}\right)\pi,\nonumber
\\A_{3}&=& \frac{2}{3} \left(35a_{1}^3 + 15a_{1}^2b_{1}-3a_{1}(10a_{2}+b_{1}^2-4b_{2})+ 6a_{3}\nonumber
\right.\\&+&\left. b_{1}^3 - 6a_{2}b_{1}-4b_{1}b_{2}+8b_{3}\right).\label{MR1}
\end{eqnarray}
The spacetime metric for WH-like static aether solution already defined by Eq.$(\ref{AH0})$;
\begin{equation}
    ds^2=-A(r) dt^2+B(r) dr^2+D(r) d\Omega^2,\nonumber
\end{equation}
\begin{equation}
\text{with}~~~~d\Omega^{2}=d\theta^2+\sin^2\theta d\phi^2,~~~~~~~~~~~~~~~~~~~~~~~~~~~~~~~~~~\nonumber\\
\end{equation}
\begin{equation}
A(r)=\left({\frac{1-\frac{m}{2r}}{1+\frac{m}{2r}}}\right)^q,~~~~~B(r)=\frac{({1+\frac{m}{2r}})^{q+2}}{({1-\frac{m}{2r}})^{q-2}}. \nonumber\\
\end{equation}
Dividing the right hand side of the metric with $B(r)$, where $B(r)=D(r)/r^2$.
Then the standard form of metric is written as;
\begin{eqnarray}
ds^2=-\frac{A(r)}{B(r)}dt^2+dr^2+r^2d\Omega^2, \label{sib22}
\end{eqnarray}
\begin{equation}
\text{where}~~~~~~~~G(r)=\frac{A(r)}{B(r)}=1-\frac{2qm}{r}+\frac{(1+4q^2)m^2}{2r^2}+\frac{(-7q-8q^3)m^3}{6r^3}+\mathcal{O}\left({m^4}\right).\label{sib24}
\end{equation}
\begin{equation}
\text{and}~~~~~~~~H(r)=1,~~~~~~~~~~~~~\nonumber
\end{equation}
Now, we compare the $G(r)$ with  $\bar{A}(r)$ and  $\bar{B}(r)$ with $H(r)$ and write the PPN coefficients as;
\begin{eqnarray}
~~a_1=q,~~~~a_2=q^2+\frac{1}{4},~~~~a_3=\frac{7q+8q^3}{12}, ~~~~~~~b_1=b_2=b_3=0. \nonumber
\end{eqnarray}
After putting all above coefficient into Eq.(\ref{MR1}) we get;
\begin{eqnarray}
A_1=2q,~~~~A_2=\left(q^2-\frac{1}{4}\right)\pi,~~~~~~A_3=6q^3-\frac{8q}{3}. \nonumber
\end{eqnarray}
Hence, the bending angle for WH-like static aether solution by Keeton and Petters method can be computed as;
\begin{eqnarray}
\psi=2q\left(\frac{m}{b}\right)+\left(q^2-\frac{1}{4}\right)\pi\left(\frac{m}{b}\right)^2+\left(6q^3-\frac{8q}{3}\right)\left(\frac{m}{b}\right)^3+\mathcal{O}\left(\frac{m}{b}\right)^4. \label{sib28}
\end{eqnarray}
The obtained deflection angle depends on $m$, $q$ and  $b$. The obtained angle (\ref{sib28}) reduces to the deflection angle of Schwarzschild by using Keeton and Petters technique when $q=2$.

\section{Photon ring and wormhole shadow}
In this section, let us examine the photonsphere and the shadow produced by the wormhole considered in this study. There have been various studies of the shadow of black holes and shadow of wormholes \cite{Kuang:2022xjp,Uniyal:2022vdu,Khodadi:2020jij,Vagnozzi:2022moj,Roy:2021uye,Vagnozzi:2019apd,Allahyari:2019jqz,Atamurotov:2013sca,Abdujabbarov:2015xqa,Wei:2019pjf,Wei:2018xks,Abdolrahimi:2015rua,Adair:2020vso,Abdolrahimi:2015kma,Herdeiro:2021lwl,Cunha:2019hzj,Cunha:2019ikd,Cunha:2018acu,Cunha:2016wzk,Afrin:2021imp,Jha:2021bue,Khodadi:2021gbc}. But first time here we will include the influence of a non-magnetized cold plasma with electron plasma frequency $\omega_e(r)$,  which can be done through by means of deriving the equations of motion (EoS) through the Hamiltonian \cite{Perlick2015} to wormhole spacetime.
\begin{equation}
    H = \frac{1}{2} g^{ik} p_{i} p_{k} = \frac{1}{2} \left( -\frac{p_{t}^{2}}{A(r)} + \frac{p_{r}^{2}}{B(r)} + \frac{p_{\phi }^{2}}{C(r)} +\omega_e(r)^2\right).
\end{equation}
We only considered motion along the equatorial plane, thus, $D(r)=C(r)$. We can then derive the EoS through
\begin{equation}
    \dot{x}^{i} = \frac{\partial H}{\partial p_{i}}, \quad \quad \dot{p}_{i} = -\frac{\partial H}{\partial x^{i}},
\end{equation}
which enables us to extract the two constants of motion:
\begin{equation}
    E = A(r)\frac{dt}{d\lambda}, \quad L = C(r)\frac{d\phi}{d\lambda}.
\end{equation}
Also, using this, we can define the impact parameter as
\begin{equation}
    b \equiv \frac{L}{E} = \frac{C(r)}{A(r)}\frac{d\phi}{dt}.
\end{equation}
Going back to the metric, null geodesics requires that $ds^2=0$, and the orbit equation can then be expressed as
\begin{align}
    \left(\frac{dr}{d\phi}\right)^2 =\frac{C(r)}{B(r)}\left(\frac{h(r)^2}{b^2}-1\right).
\end{align}
Following methods in Ref. \cite{Perlick2015}, the orbit equation allows one to define the function
\begin{equation}
    h(r)^2 = \frac{C(r)}{A(r)}n(r)^2=\frac{C(r)}{A(r)}\left(1-\frac{\omega_e^2}{\omega_0^2}A(r) \right),
\end{equation}
under the assumption that the homogeneous plasma is non-gravitating \cite{crisnejo2018weak:2018xnr}. It is also easy to see how the above reduces to the standard case if $n(r)=0$. The photonsphere can then be sought off by solving $r$ in $h'(r)=0$. Depending how complicated the expression for the metric coefficients whether one can obtain an analytic expression or not. One can determine the photonsphere radii via
\begin{equation}
    \left(\frac{\omega_{e}^{2}}{\omega_{0}^{2}}A(r)^{2}-A(r)\right)C'(r)+C(r)A'(r)=0,
\end{equation}
and for the case of $n(r)=0$, we get
\begin{equation} \label{erph}
    r_\text{ph}=\frac{m}{2}\omega^{\pm},
\end{equation}
where we write first $\omega^{\pm} = q\pm \sqrt{q^2-1}$ for brevity \cite{epps2017weak:2017xnr}. We note that there is a third solution $r_\text{ph}=m/2$, but it does not produce any shadow cast due to the wormhole.

\begin{figure}
   \centering
    \includegraphics[width=0.50\textwidth]{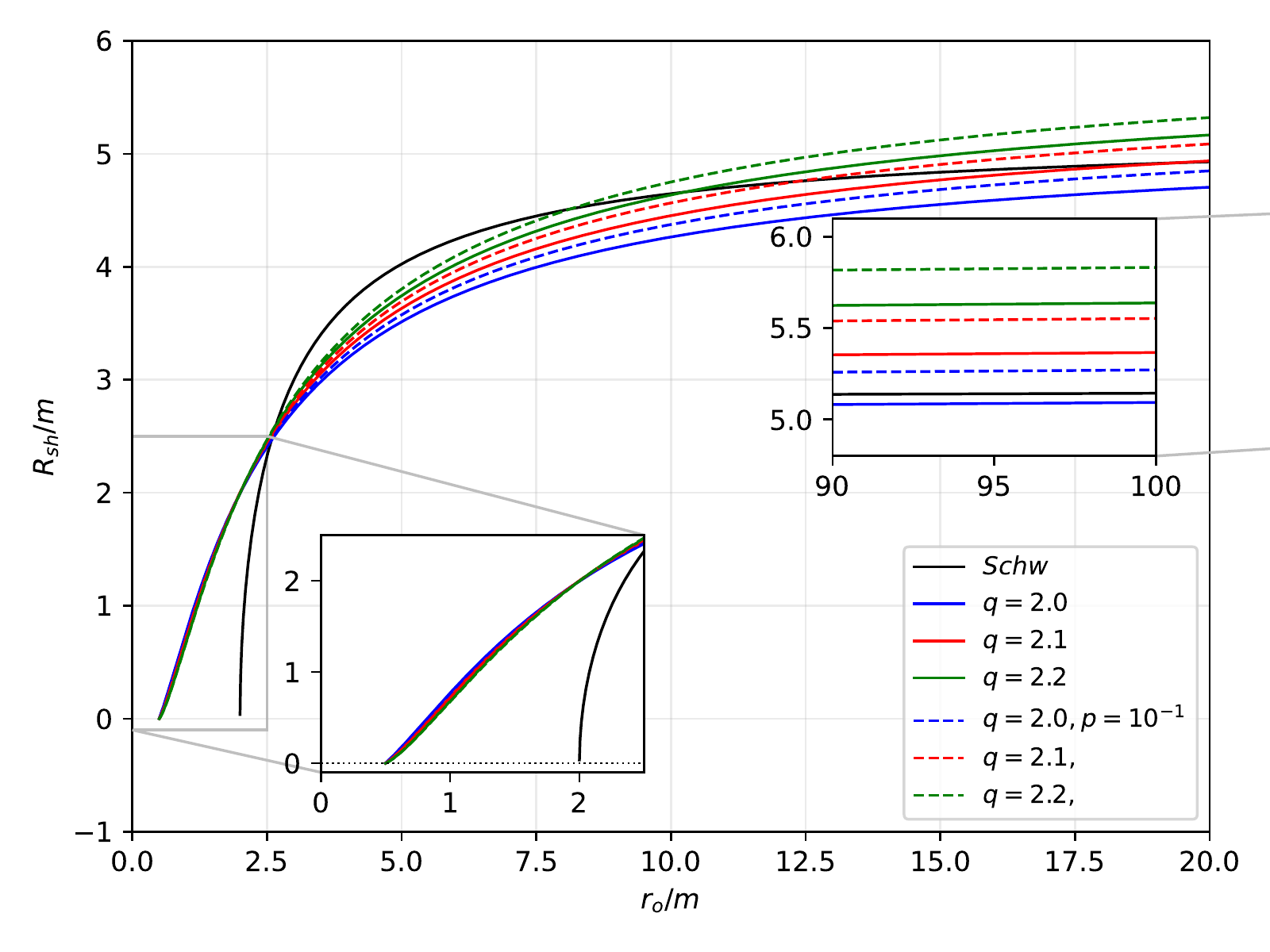}
    \caption{Plot of the shadow radius of the wormhole at varying locations of a static observer. Here, we compared the shadow behavior between the Schwarzschild case, and the wormhole with and without the influence of plasma. We set $m=1$ and the plasma parameter $p=10^{-1}$.}
    \label{sharad}
\end{figure}

A static observer at infinity can construct the following relation,
\begin{equation}
    \tan(\alpha_{\text{sh}}) = \lim_{\Delta x \to 0}\frac{\Delta y}{\Delta x} = \left(\frac{C(r)}{B(r)}\right)^{1/2} \frac{d\phi}{dr} \bigg|_{r=r_\text{obs}},
\end{equation}
which can be simplified into
\begin{equation} \label{eang}
    \sin^{2}(\alpha_\text{sh}) = \frac{b_\text{crit}^{2}}{h(r_\text{obs})^{2}}
\end{equation}
with the help of the orbit equation. The critical impact parameter can be sought off under the condition $d^2r/d\phi^2 = 0$ and we find
\begin{equation}
    b_\text{crit}^2 = \frac{h(r_\text{ph})}{\left[B'(r_\text{ph})C(r_\text{ph})-B(r_\text{ph})C'(r_\text{ph})\right]} \Bigg[h(r_\text{ph})B'(r_\text{ph})C(r_\text{ph})-h(r_\text{ph})B(r_\text{ph})C'(r_\text{ph})-2h'(r_\text{ph})B(r_\text{ph})C(r_\text{ph}) \Bigg],
\end{equation}
where the derivatives with respect to $r$ are evaluate at $r \to r_\text{ph}$. The analytic expression is quite lengthy with the inclusion of plasma, but for the case without its influence, we can obtained two solutions:
\begin{equation}
    b_\text{crit}^2=\frac{m^{2}\sqrt{q^{2}-1}(\omega^\pm \mp 1)^{1-2q}(\omega^\pm \pm 1)^{2q+1}}{2\omega^\pm}.
\end{equation}
This will be used to the calculation of the shadow, which gives us the exact analytical formula of
\begin{equation} \label{eshad}
    R_\text{sh}=\left[\frac{8r_\text{obs}^{4}m^{2}\sqrt{q^{2}-1}(\omega^{\pm}\mp1)^{2q+1}(\omega^{\pm}\pm1)^{1-2q}(2r_\text{obs}-m)^{2(q-1)}(2r_\text{obs}+m)^{-2(q+1)}}{\omega^\pm}\right]^{1/2}
\end{equation}
for the case $n(r)=0$. For the case with plasma, we plot it numerically. See Fig. \ref{sharad}.
We can then see how the shadow radius behave in relation to the location of the static observer with respect to the wormhole. We can see that the Schwarzschild case behave as it is. But, when the wormhole is considered, we can see that the zero radius shadow is nearer to the black hole. The lower left inset plot reveals that the effect of plasma is not that evident. But we can see that the shadow radius of the wormhole slightly increases, then decreases again. The intersection point, near $r=2m$, indicates that the angular radius of the shadow is $\theta=\pi/2$. In this location, the observer will see that half of the sky is dark. After, we can see the obvious deviations at farther distances. Notably, the plasma's effect is to increase the wormhole shadow. At very large distances, we can see that the rate of the change in the shadow radius levels-off near the Schwarzschild case. It indicates that the shadow can be useful to detect the imprints of plasma. Finally, we remark that in this plot, we have only used the upper sign in Eq. \eqref{sharad}. Using the lower sign, one can verify that using $q>0$ does not produce any shadow, and using $q<0$ gives infinitely large shadow near the wormhole, which is unphysical. However, we found out that at very large distances, the effect of the second solution is nearly the same as the one in the upper right inset plot.

Let us now use the DM refractive index $n(\omega)$ in Eq. \eqref{sib8} for the next case. We then find that the location of the photonsphere is independent of $n(\omega)$:
\begin{equation}
    n(\omega)^{2}[C'(r)A(r)-A'(r)C(r)]=0,
\end{equation}
which yields the same expression for the photonsphere in Eq. \eqref{erph}. For the critical impact parameter, we find
\begin{equation}
    b_\text{crit}^2 =\pm\frac{n(\omega)^{2}m^{2}\sqrt{q^{2}-1}(\omega^{\pm}\mp1)^{2(1-q)}(\omega^{\pm}\pm1)^{2(q+1)}}{4\omega^{\pm}(\omega^{\pm}q-1)},
\end{equation}
where we are only interested in using the upper sign. With Eq. \eqref{eang}, we can get an analytical expression for the shadow radius as
\begin{equation}
    R_\text{sh}=4mn(\omega)(-1)^{q}r_\text{obs}^{2}(m-2r_\text{obs})^{q-1}(m+2r_\text{obs})^{-(q+1)}\left[\pm\frac{\sqrt{q^{2}-1}(\omega^{\pm}\mp1)^{2(1-q)}(\omega^{\pm}\pm1)^{2(q+1)}}{4\omega^{\pm}(\omega^{\pm}q-1)}\right]^{1/2},
\end{equation}
which is quite a worked out equation. Interestingly, for static observers in a remote location from the wormhole, we can apply Taylor expansion to get a simplified and approximated equation:
\begin{equation}
    R_\text{sh} = n(\omega)m\left[\pm\frac{\sqrt{q^{2}-1}(\omega^{\pm}\mp1)^{2(1-q)}(\omega^{\pm}\pm1)^{2(q+1)}}{4\omega^{\pm}(\omega^{\pm}q-1)}\right]^{1/2}.
\end{equation}
In a case where $q=2$, we find
\begin{equation}
    R_\text{sh}=3\sqrt{3}mn(\omega),
\end{equation}
where we could see clearly the influence of dark matter to the shadow radius. Furthermore, in this remote region, we saw again that the effect of the wormhole mimics the Schwarzschild case.

\section{Conclusion}

In this paper, we have discussed WH-like static aether solution and derived deflection angle in the non-plasma, plasma and DM mediums. Also, we have found the deflection angle by using Keeton and Petters technique. For this purpose, we have used an optical metric to determine Gaussian optical curvature and then applied GBT to examine the deflection angle.

We have examined that deflection angle  $(\ref{P1})$ in the non-plasma medium, plasma medium $(\ref{P2})$, DM medium $(\ref{sib14})$ and by Keeton and Petters technique $(\ref{sib28})$ is depending upon the parameter $q$, mass $m$ as a integrating constant and impact parameter $b$.

The graphical behaviour of bending angle in the plasma medium has examined in such a way that when we make a relation between $\psi$ and $b$ and vary the value of $q$. We have seen that the deflection angle increases when the values of $q$ increases. Moreover, when we make a relation between $\psi$ and $m$ and varying $b$, we have noticed that the angle is decrease. Afterwards, we have noticed that when we make a relation between $\psi$ and $q$ and varying $m$, the angle is increases. It is to be mentioned here that the plots in non-plasma have shown the same behaviour as plasma medium.

We have observed that if the plasma effect is ignored as $(\frac{\omega_e}{\omega_\infty}\rightarrow0)$, then the bending angle $(\ref{P2})$ has reduced into the bending angle $(\ref{P1})$. In case of DM medium, we have observed that if we removed the effect of DM medium then this obtained angle converts into the angle obtained in $(\ref{P1})$.
We have also examined that the obtained deflection angles by plasma, non-plasma, DM and Keeton and Petters technique reduces to the Schwarzschild deflection angle upto $1$st order term of $m$ by taking $q=2$.

The results we have obtained for WH-like static aether solution in the presence of different mediums i.e plasma and non-plasma shows that deflection angle $\psi$ has direct relation with mass $m$ and a parameter $q$ which means that WH with greater mass has greater gravitational pull and bends the light passing by it at large angle. Whereas, WH with smaller mass deflect the light at smaller angle. We also notice that deflection angle $\psi$ has inverse relation with impact parameter $b$, which shows that smaller value of impact parameter has larger deflection angle and vice versa.

Also, we have examined the impact of DM medium on bending angle of WH-like static aether solution. The refractive index in DM medium has taken homogeneously nonuniform. Hence it is concluded that bending angle by WH-like static aether solution increases with increasing parameter $q$ and $m$, while the bending angle decreases in a increasing medium of DM. It is showed that how weak deflection angle of WH is the affected by parameter $q$ and $m$.

To broaden the scope of the study, we also examined the behavior of the shadow radius of the wormhole, comparing it to the case where it is surrounded by plasma. Our main result indicates that as the photons travels through the plasma, its imprints or effects can be perceived by a static observer at infinity through the increased shadow size. Although highly unlikely in terms of situational applicability, our calculation also reveals that for observers near the wormhole, the effects of plasma is rather weak, compared to an observer at a very large distance.

\acknowledgements
 A. {\"O}. and R. P. would like to acknowledge networking support by the COST Action CA18108 - Quantum gravity phenomenology in the multi-messenger approach (QG-MM).

\bibliography{ref}
\bibliographystyle{apsrev}

\end{document}